\documentclass[10pt,aps,prl,twocolumn,amsmath,amsonts,showpacs,floatfix]{revtex4-1}
\usepackage{graphics}
\usepackage{graphicx}

\begin{document}

\title{Hippocampal Spike-Timing Correlations Lead to Hexagonal Grid Fields}

\author{Mauro M Monsalve-Mercado, Christian Leibold} 

\affiliation{Department Biologie II \& Graduate School of Systemic
  Neurosciences, LMU Munich, Gro{\ss}hadernerstr. 2, 82152 Planegg,
  Germany}  
\affiliation{Bernstein Center for Computational
  Neuroscience Munich, Gro{\ss}hadernerstr. 2, 82152 Planegg, Germany}

\date{\today}

\begin{abstract}
Space is represented in the mammalian brain by the activity of
hippocampal place cells as well as in their spike-timing
correlations. Here we propose a theory how this temporal code is
transformed to spatial firing rate patterns via spike-timing-dependent
synaptic plasticity. The resulting dynamics of synaptic weights
resembles well-known pattern formation models in which a lateral
inhibition mechanism gives rise to a Turing instability. We identify
parameter regimes in which hexagonal firing patterns develop as they
have been found in medial entorhinal cortex.
\end{abstract}

\maketitle

%\section{Introduction}
The spatial position of an animal can be reliably decoded from the
neuronal activity of several cell populations in the hippocampal
formation~\cite{harris,mathis,kammerer14}. For example, place cells in
the hippocampus fire at only few locations in a spatial
environment~\cite{Okeefe,muller} and the position of the animal can be
readily read out from single active neurons. Grid cells of the medial
entorhinal cortex (MEC) fire at multiple distinct places that are
arranged on a hexagonal lattice~\cite{fyhn,hafting}. Although
hexagonal patterns are abundant in nature and there exist well-studied
physical theories for their emergence, the mechanistic origin of this
neuronal grid pattern is still unclear.  Initially it was suggested
that they result from continuous attractor dynamics~\cite{fuhs,burak}
or superposition of plane wave inputs~\cite{burgess} and, based on
circuit anatomy, place cells would then result from a superposition of
many grid cells~\cite{mcnaughton_moser06_NRN,moser08}.  More recent
experiments, however, reported place cell activity without intact grid
cells, such that grid cells are not the unique determinants of place
field firing~\cite{koenig,brandon,wills,hales,schlesiger}. Conversely,
it would thus be possible that grid fields may arise from place field
input as suggested in~\cite{kropff,derdikman,stepanyuk}. The
biological mechanisms proposed by these latter theories, however,
remain hypothetical. In the present Letter, we propose a learning rule
for grid cells based on the individual spike timings of place cells
using spike-timing dependent synaptic plasticity
(STDP)~\cite{markram,kempter,bi}. The theory thereby predicts that the
observed temporal hippocampal firing patterns (phase precession and
theta-scale correlations; see
below)~\cite{OKeefeRecce,dragoi06,foster} translate the temporal
proximity of sequential place field spikes into spatial
neighborhood-relations observed in grid-field activity. For our model
to work, we only have to assume that the synaptic plasticity rule
averages over a sufficiently long time interval.

\paragraph*{Model.}
%\section{Model}
We use the classical formulation of pairwise additive
STDP~\cite{gerstner,kempter}, where the update of a synaptic weight
$J_n\ ,\ n=1,\ldots, N$ at time $t$ is computed as~\cite{kempter}
\begin{equation}\label{eq-stdpmeanfield}
\frac{\rm d}{{\rm d}t} J_n = \int_{-\infty}^{\infty}{\rm d}s\,
W(s)\, C_n(s) + F(J_n)\ .
\end{equation}
$C_n(s)$ denotes the time averaged correlation function between the
spike train of presynaptic neuron $n$ and the postsynaptic neuron, the
learning window $W(s)$ describes the update of the synaptic weight as
a function of the time difference $s$ between a pair of pre- and
postsynaptic action potentials, and the function $F$ implements soft
bounds for the weight increase.  The dynamics is further constrained
such that weights cannot become negative.

%\paragraph*{Neuron model.}
To be able to treat eq.~(\ref{eq-stdpmeanfield}) analytically, we use
a linear Poisson neuron model, i.e., the mean firing rate of the
postsynaptic neuron $E(t) = \mathbf{J}\cdot \mathbf{H}(t)$ results
from a weighted sum of hippocampal firing rates $\mathbf{H} = (H_1(t),
\ldots, H_N(t))^{\rm T}$. Under these assumptions $C_n(s)$ can be
approximated for large $N$~\cite{kempter} as $C_n(s) = \sum_{n'}
J_{n'} C_{nn'}(s)$ with
\begin{equation}\label{eq-CPoisson}
C_{nn'}(s) := \int_{-\infty}^{\infty} {\rm d}t\, H_n(t)\, H_{n'}(t-s)\ .
\end{equation}
Inserting the correlation functions from eq.~(\ref{eq-CPoisson}) into
the weight dynamics from eq.~(\ref{eq-stdpmeanfield}) yields
\begin{equation}\label{eq-learningequation}
\frac{\rm d}{{\rm d}t} J_n \!=\! \sum_{n'} J_{n'} G_{nn'}\! +\!
F(J_n)\, , \
G_{nn'} \!:=\!\!\! \int_{-\infty}^{\infty}\!\!\!\!\!\!
{\rm d}s\, W(s)\, C_{nn'}(s)\, .
\end{equation}
Following \cite{vanrossum,kistler} we introduce the quadratic
stabilization term $F(J) = F_0\, J\, (K - J)\ , F_0>0$ that
implements a soft upper bound.

%\paragraph*{Hippocampal Phase Precession.}
As an input to the postsynaptic neuron, we consider a population of
$N$ hippocampal place cells. The firing of these neurons is
characterized by a bell shaped envelope modulating the spatial path
$x_{\cal P}(t)$ and oscillations in time $t$ (Fig.~1A), 
\begin{equation}\label{eq-pp}
 H_n(t;{\cal P}) = a\, {\rm e}^{\frac{-(x_{\cal P}(t)-x_n)^2}{2\sigma^2}}\,
 [\cos(\omega\, t+\phi_n) + 1]/2\ .
\end{equation}
The oscillation frequency $\omega$ of a neuron is slightly higher
than the frequency $\omega_\theta$ of the theta oscillation ($\sim
8$~Hz) in the local field potential giving rise to a phenomenon called
theta phase precession (Fig.~1B): spikes early in the field come at
later phases than spikes late in the field~\cite{OKeefeRecce}. 
During traversal of a place field, phase precession spans a whole
theta cycle~\cite{maurer}. Thus the two frequencies have to relate to
each other like $\omega =\omega_\theta + \frac{\pi}{R}\,v$, with $R$
denoting the distance from the place field center at which the firing
rate has decreased to $10\%$ (i.e., $R=\sigma\, \sqrt{2\, \ln(10)}$),
and $v$ denoting the running speed, which we fix at 25~cm/s. At each
individual entry into the place field, the phase of the cellular
oscillation is reset to phase zero with respect to the theta
oscillation phase $\phi_\theta$ by fixing $\phi_n=(\phi_\theta -
2\pi)\, \omega/\omega_\theta$.

To obtain a closed expression for the correlation function
$C_{nn'}(s)$ (Fig.~1C), the time average in eq.~(\ref{eq-CPoisson}) is
performed over all straight paths ${\cal P}$ crossing the center of
the field overlap. For place fields with identical width $R$, firing
rates $a$, and at small lags $s$, we obtain
\begin{eqnarray}\label{eq-Cpp}
 C_{nn'}(s)&=& \int_{\cal P}  \int_{-\infty}^{\infty}\!\!\!\! {\rm d}t\, H_n(t;{\cal P})\,
 H_{n'}(t-s;{\cal P})\\ &=&a\, \frac{\sqrt{\pi}\, {\rm
     e}^{-\frac{r^2 + v^2\, s^2}{4\sigma^2}}}{4v/(a \sigma)}\,  \Bigl(1
 + \frac{1}{2}J_0\left(\frac{\pi r}{R}\right)\, \cos(\omega s)\nonumber\\
& & \qquad\qquad\qquad +\ \frac{v s r}{4\sigma^2} \, J_1\left(\frac{\pi
   r}{R}\right)\, \sin(\omega s)\Bigr)\ ,\ \nonumber
\end{eqnarray}
with place field distance $r=|x_n-x_{n'}|$, and $J_{0,1}$ denoting
Bessel functions of first kind (see Supplementary Material at [URL]
for derivation).

\begin{figure}[bt]
\includegraphics[width=\columnwidth]{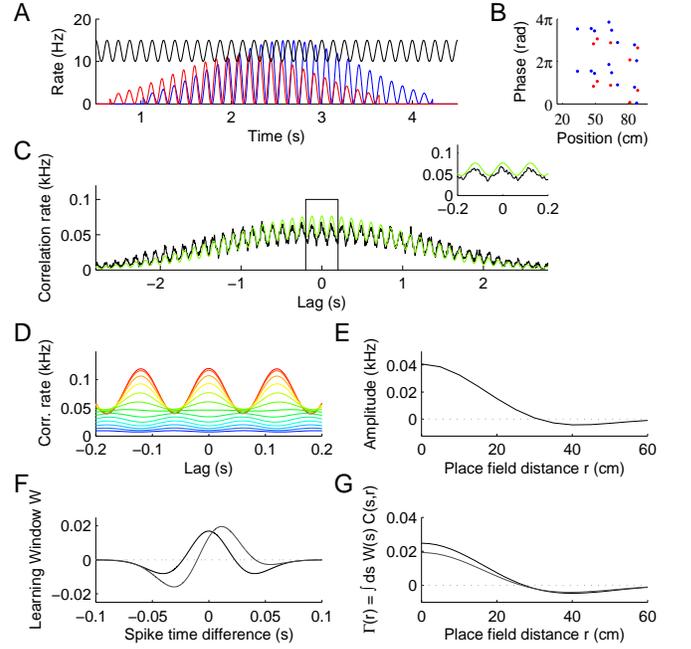}
\caption{(A) Poisson model of place field firing for two place cells
  (red, blue), and slower theta oscillation (black). (B) Phase
  precession resulting from the path used in A. (C) Correlation of
  spike trains for the cells in A averaged over 2-d trajectories
  (random walk) (black), and from eq.~(\ref{eq-Cpp}) (green). Inset
  magnifies lag 0. (D) Correlation functions for different place field
  distances (red: autocorrelation; blue: 60~cm). (E) Correlation
  amplitude as function of place field distance. (F) Examples of STDP
  learning windows; see eq.~(\ref{eq-W}) and below. (G) $\Gamma$
  kernels for correlation function from C-E and learning windows from
  F.}\label{fig1}
\end{figure}

In contrast to 1-d fields where the distance of place fields is
reflected by the lag of the correlation peak (theta
compression)~\cite{dragoi06}, correlation functions in 2-d are
symmetric because of the symmetry of the path, however, the distance
of the place field centers is encoded in the amplitude of the
correlation peak at lag 0 (Fig.~\ref{fig1}D,E).

\paragraph*{Weight Dynamics.}
Assuming that the putative grid cells receive inputs from a large
number $N\gg 1$ of place cells that sufficiently cover the encoded
area, we replace the presynaptic index $n$ by the position of the
place field center $x$, i.e., $G_{nn'} \to \Gamma(|x_n-x_{n'}|)$,
and thereby translate the learning eq.~(\ref{eq-learningequation}) to
continuous coordinates,
\begin{eqnarray}\label{eq-learningequationcont}
\frac{\rm d}{{\rm d}t} J(x) &=&  (\Gamma\ast J)(x) + F_0 J(x)\,(K - J).
\end{eqnarray}
Examples of the convolution kernel $\Gamma(|x|)$ for different
learning window functions $W$ are depicted in Fig.~\ref{fig1}F,~G.
The development of the weights follows the pattern formation
principles of a lateral inhibition system~\cite{Swindale80}. Indeed,
the integro-differential equation
(IDE)~(\ref{eq-learningequationcont}) involves non-local interactions
effectively implemented through the convolution kernel, inducing a
strong close range potentiation and a weaker long range depression of
neighbouring synapses, as observed in the typical shape presented in
Fig.~\ref{fig1}G. A general window-dependent kernel can be obtained
for the correlation function from eq.~(\ref{eq-Cpp}) as
\begin{equation}\label{Gamma}
\Gamma(r) = c\,\frac{\sqrt{\pi} {\rm
    e}^{-\frac{r^2}{4\sigma^2}}}{4v/(a\,\sigma)}\left(1+\alpha J_0\left(\frac{\pi r}{R}\right) +
\frac{\beta r}{\sigma}J_1\left(\frac{\pi r}{R}\right) \right)
\end{equation}
\begin{eqnarray}\label{functionals}
 c[W]: &=&
 a\int_{-\infty}^{\infty}\exp[-\frac{v^2s^2}{4\sigma^2}]\,W(s)\,{\rm d}s
 \\ \alpha[W]: &=& \frac{a}{2c}
 \int_{-\infty}^{\infty}\exp[-\frac{v^2s^2}{4\sigma^2}]\,\cos(\omega
 s)\,W(s)\,{\rm d}s \nonumber\\ \beta[W]: &=& \frac{av}{4\sigma c}
 \int_{-\infty}^{\infty}\exp[-\frac{v^2s^2}{4\sigma^2}]\,s\,\sin(\omega
 s)\,W(s)\,{\rm d}s \nonumber
\end{eqnarray}
which can take a Mexican-hat type shape for qualitatively different
learning window functions $W$, due to the symmetry of the correlation
function (Fig.~\ref{fig1}F,~G). To see this, we can regard
Hebbian-like windows to be modelled as the product of a Gaussian and a
polynomial of some order $m$,
$W(s)=\exp[-s^2/(2\rho^2\mu^2)]\, P_m(s/\rho)$.  The functionals defined
in eq.~(\ref{functionals}) then inherit the symmetries from the
cross-correlation, since all of the odd terms in the polynomial cancel
out during integration. Thus, in subsequent numerical investigations
we focus on windows up to second polynomial order
\begin{equation}\label{eq-W}
 W(s)= W_0\,(2\pi\mu^2\rho^2)^{-\frac{1}{2}} \left(1-(s/\rho)^2\right)\, {\rm
   e}^{-\frac{s^2}{2\rho^2\mu^2}},
\end{equation}
whose free parameters $\rho$ and $\mu$ determine its zeroes $s_0=\pm
\rho$ and negativity $\int W=W_0(1-\mu^2)$. In Figure~\ref{fig1}F,G we
used $\rho=23$~ms, $\mu=1.025$ and added a linear term $s/\rho$ to the
polynomial to get the asymmetric window (grey lines).  The
$W$-dependent functions $c, \alpha$, and $\beta$ defining the kernel
$\Gamma$ are given in the Supplementary Material [URL].

\begin{figure}[b]
\includegraphics[width=\columnwidth]{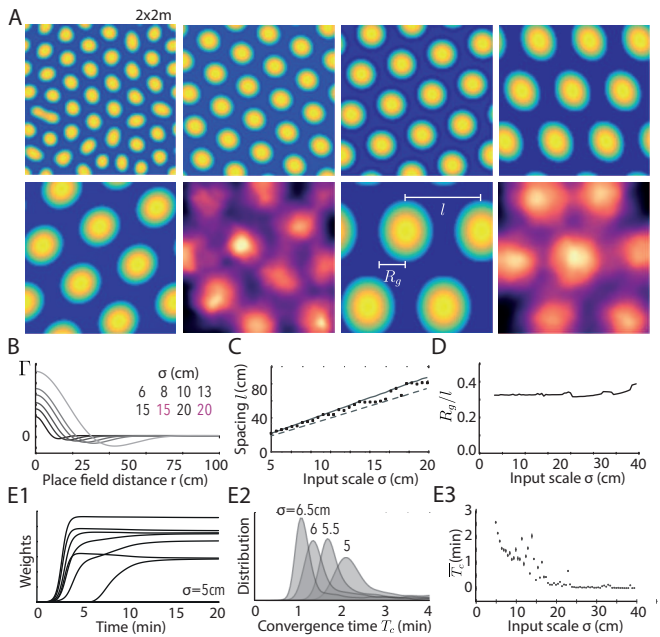}
\caption{Weight patterns and input scale. (A) Asymptotically stable
  weight functions $J(x)$ for place field widths $\sigma$ as indicated
  in (B) together with the respective $\Gamma$ kernels ($\mu=1.025$,
  $\rho=23$~ms, $\sigma$ indicated by grey level). Firing rates of
  spiking simulations are colored in red (for numerical details see
  Supplementary Material [URL]). (C) Grid spacing $l$ (see A) scales
  linearly with place field scale $\sigma$. Dots correspond to weight
  patterns from numerical solutions. Dashed and solid lines indicate
  theoretical estimates $2\pi/k_m$ and $2\pi/\tilde{k}_m$; see {\em
    Linear theory at early times} and {\em Positivity constraint.} (D)
  Estimated ratio of field radius $R_g$ (see A) to grid spacing $l$,
  $R_g/l=(\sqrt{3}\nu/(2\pi))^{1/2}$ is independent of
  $\sigma$. (E$_1$) Temporal evolution of randomly sampled weights.
  (E$_2$) Distribution of weight convergence times $T_c$ for different
  input scales $\sigma$ as indicated and the respective means
  (E$_3$).}\label{fig2}
\end{figure}

A numerical evaluation of the learning
IDE~(\ref{eq-learningequationcont}) with periodic boundary conditions
reveals that the spatially isotropic kernel $\Gamma$ can result in
hexagonal packing structures~(Fig.~\ref{fig2}A,~B). Simulations of
spiking Poisson neurons confirm these predictions of the meanfield
theory (Fig.~\ref{fig2}A). As indicated by the kernel function in
eq.~(\ref{Gamma}), the grid spacing only depends on the spatial scale
$\sigma$ of the place fields in the input (Fig.~\ref{fig2}C).

An experimentally accessible quantity to compare our model results to
is the ratio of grid field radius $R_g$ to grid spacing $l$ as
indicated in Fig.~\ref{fig2}A. In experiments, the ratio $R_g/l$ for
grid cells has been determined to be about $0.3$~\citep{hafting}, and
for a perfectly hexagonal grid $R_g/l$ relates to the fraction $\nu$
of field size per area as $\nu=\pi\, R_g^2/(\frac{\sqrt{3}}{2}\,
l^2)$. The field fraction $\nu$ can be readily accessed from our
numerics as the fraction of non-zero synaptic weights, and for the
used learning window fits the experimentally obtained $R_g/l$
(Fig.~\ref{fig2}D) for all choices of $\sigma$. Finally, learning
converges faster for large spacing (large $\sigma$) consistent with a
larger amplitude of $\Gamma$ [eq.~(\ref{Gamma}) and
  Fig.~\ref{fig2}B,E].

\paragraph*{Linear theory at early times.}
Some analytical understanding of the weight dynamics from
eq.~(\ref{eq-learningequationcont}) can be gained from a neural field
theory approach
\cite{Swindale80,Linsker86,Ermentrout86,Miller89,Shouval96}. In this
framework, we can neglect the effect of the non-linearities at early
times, and focus only on the convolution term $\Gamma\ast J$. The
emerging dynamics can be readily understood by looking at the
evolution of the weights in Fourier space
$\partial_t\hat{J}(k)=\hat{\Gamma}(k)\, \hat{J}(k)$, which makes
evident that the wave number $k_m$ maximizing the kernel Fourier
transform $\hat{\Gamma}$ (see Supplementary Material [URL]) will
exponentially overgrow all other modes (if $\hat{\Gamma}(k_m){>}0$),
thus setting the initial periodicity of the pattern.

\begin{figure}[b]
\includegraphics[width=\columnwidth]{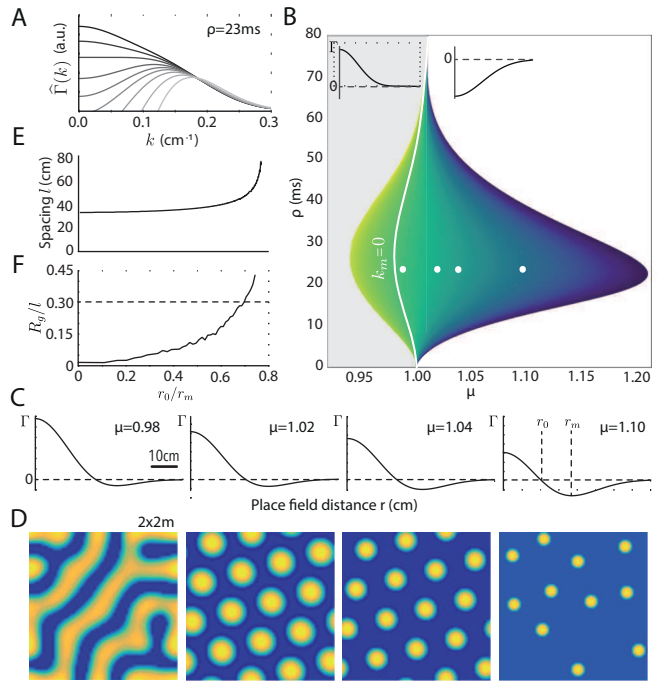}
\caption{Permitted learning windows. (A) Fourier transforms of
  $\Gamma$ for $\mu=0.95$ (black) to $1.20$ (light grey). (B) Region
  of structure formation in ($\mu, \rho$) space (for
  $\sigma=10$~cm). Non-trivial patterns appear to the right of the
  white solid line ($k_m=0$), where the selected wavelength is
  positive. Colored region indicates bimodal $\Gamma$. The color
  encodes the shape factor $r_0/r_m$ (0: blue; 1: yellow). Grey and
  white areas correspond to regions where $\Gamma$ is all positive or
  all negative. (C) Four examples of $\Gamma$ kernels corresponding to
  the white dots from B in the same order. (D) Weight patterns for kernels
  from C.  (E) Predicted spacing $2\pi/k_m$ as a function of $r_0/r_m$
  obtained for all combinations of $\mu$ and $\rho$ from the region
  right of the solid line in A. (F) Estimated ratio of field size to
  grid spacing as a function of $r_0/r_m$ (from simulated
  patterns). Dashed line indicates experimental value~\cite{hafting}.
}\label{fig3}
\end{figure}

In a finite region of the parameter space $(\mu, \rho)$ of the
learning window, the bimodal (Mexican-hat) shape of $\Gamma$ ensures
the existence of a Turing instability, i.e., a transition to single
maximum of $\hat{\Gamma}$ (with $\hat{\Gamma}(k_m){>}0$) at a non-zero
$k_m{>}0$ (Fig.~\ref{fig3}A-C).  Similar to previous work on pattern
formation in lateral inhibition systems~(e.g. \cite{ermentrout79}),
the permitted parameter region ($k_m$ exists and is positive) gives
rise to stripe-like and hexagonal patterns (Fig.~\ref{fig3}D). In the
Supplementary Material [URL], we also provide a complementary
description of the pattern formation process based on an approximation
of eq.~(\ref{eq-learningequationcont}) by a partial differential
equation (Swift-Hohenberg equation~\cite{swifthohenberg}), which
corroborates the results from the linear theory.

To connect the resulting patterns to other feed-forward models of grid
field formation, we parameterize $\Gamma$ by the shape factor
$r_0/r_m$ (Fig.~\ref{fig3}C), which is the fraction between the zero
and the minimum of $\Gamma$. The shape factor $r_0/r_m$ reduces the
two-parameter learning window to a single qualitatively descriptive
parameter, which can be used to describe the bimodal kernel $\Gamma$
independently of the hypothesized biological mechanism.  If $r_0/r_m$
is large ($\sim 0.8$), $\Gamma$ shows only little negativity and the
emerging pattern is stripe-like (Fig.~\ref{fig3}C,~D), if $r_0/r_m$ is
small, $\Gamma$ exhibits strong negativity, the firing fields become
dispersed and the pattern looses hexagonality. Hexagonal patterns
arise for $r_0/r_m$ roughly between 0.65 and 0.75
(Fig.~\ref{fig3}D). In this region, the shape factor virtually
completely determines the geometrical properties of the steady state
(Fig.~\ref{fig3}E,~F). Values of $r_0/r_m$ that give rise to hexagonal
grids can also be identified via the ratio of field width per grid
spacing $R_g/l$. According to our theory, the experimentally observed
value 0.3~\cite{hafting} is achieved with a shape factor of about
$r_0/r_m=0.7$ (Fig.~\ref{fig3}F). For higher values of $r_0/r_m$,
$R_g/l$ increases to a point where a periodic pattern cannot further
dissociate into disjoint fields and the stable pattern becomes
stripe-like. For lower $r_0/r_m$, $R_g/l$ decreases, and at some
point, the small fields no longer repel each other strongly enough to
produce a symmetrical arrangement.

\paragraph*{Positivity constraint.}  The grid spacing $l$ predicted by the
linear theory, however, consistently underestimates the spacing
derived from the numerical solution of the mean field dynamics
(Fig.~\ref{fig2}C). The reason for this error is that, after the
initial growth phase, the synaptic weights are influenced by the
non-linearities, most importantly the constraint that they cannot
become negative.

The impact of this positivity constraint can be intuitively understood
if we interpret the convolution $\Gamma\ast J$ as an operation that
detects the best overlap of an oscillatory pattern $J \propto
\cos(k\, x)$ with a given kernel $\Gamma$. However, after the
lowest weights reach zero they stop contributing to the convolution
and a slightly lower wave number $\tilde{k}_m$ maximizing
\begin{equation}\label{eq-T}
\tilde{\Gamma}(k):=\int_\Omega {\rm d}x\, \Gamma(x) \cos(k\,
x)\Theta[\cos(k\, x)-\cos(|k|\, R_g)]
\end{equation}
will be favored as the fastest growing mode ($\Theta$ denoting the
Heaviside function). Similarly, a particular field size $R_g$
maximizing $\tilde{\Gamma}$ will be selected.  In the experimentally
relevant case $|k|\, R_g=2\pi R_g/l=2\pi\times 0.3$, numerical
maximization of eq.~(\ref{eq-T}) yielded the predicted wave number
$\tilde{k}_m$ (solid line in Fig.~\ref{fig2}C), which
excellently agrees with the numerical solutions of the meanfield
dynamics.

\paragraph*{Conclusion.} For a large variety of STDP windows, the
spike-timing correlations of 2-d place cells can account for a
feed-forward learning of hexagonal grid patterns. Synaptic plasticity
thereby averages over running trajectories of tens of minutes, hence,
translating the temporal correlations into a dense code for space. Our
model thus predicts that grid cells are generated in the output
structures of the hippocampus, e.g., the deep layers of the medial
entorhinal cortex~\cite{witter} or the parasubiculum~\cite{vangroen}.
While our linear theory provides a good prediction of grid spacing as
well as for conditions that permit structure formation, determining
the boundary between hexagonal and stripe-like patterns is less
straight-forward and has to take into account the non-linearities. The
standard approach, non-linear bifurcation
analysis~\cite{ermentrout79,ermentrout80,ermentrout80a}, is difficult
because of the strong non-linearity introduced via the positivity
constraint, which strongly influences the selection of the final
pattern. Despite this drawback, our model provides a universal
framework in that it encompasses current models of grid field
formation that can be mapped to convolutions with Mexican hat-type
kernels that give rise to a Turing instability.

The authors are grateful to Andreas Herz and Anton Sirota for
discussions and Martin Stemmler for comments on the manuscript. This
work was funded by the German Research Association (DFG), Grant
No. LE2250/5-1.

\bibliography{grid}

\end{document}